# Multi-stakeholder Perspective on Responsible Artificial Intelligence and Acceptability in Education

[1]Karran, A.J., [2]Charland, P., [1]Martineau J-T., [1]Ortiz de Guinea Lopez de Arana, A., [1]Lesage, AM., [1]Senecal, S., [1]Leger, P-M.

[1]HEC Montreal, Montreal, Canada, [2]University of Quebec in Montreal, Canada

## Abstract

This study investigates the acceptability of different artificial intelligence (AI) applications in education from a multi-stakeholder perspective, including students, teachers, and parents. Acknowledging the transformative potential of AI in education, it addresses concerns related to data privacy, AI agency, transparency, explainability and the ethical deployment of AI. Through a vignette methodology, participants were presented with four scenarios where AI's agency, transparency, explainability, and privacy were manipulated. After each scenario, participants completed a survey that captured their perceptions of AI's global utility, individual usefulness, justice, confidence, risk, and intention to use each scenario's AI if available. The data collection comprising a final sample of 1198 multi-stakeholder participants was distributed through a partner institution and social media campaigns and focused on individual responses to four AI use cases. A mediation analysis of the data indicated that acceptance and trust in AI varies significantly across stakeholder groups. We found that the key mediators between high and low levels of AI's agency, transparency, and explainability, as well as the intention to use the different educational AI, included perceived global utility, justice, and confidence. The study highlights that the acceptance of AI in education is a nuanced and multifaceted issue that requires careful consideration of specific AI applications and their characteristics, in addition to the diverse stakeholders' perceptions.

## 1   Introduction

Education stands as a cornerstone of society, nurturing the minds that will ultimately shape our future (Baum et al., 2013). As we advance into the twenty-first century, exponentially developing technologies and the convergence of knowledge across disciplines are set to have a significant influence on various aspects of life (Roco et al., 2013), with education a crucial element that is both disrupted by and critical to progress (Penprase, 2018). The rise of artificial intelligence (AI), notably generative AI and generative pre-trained transformers (Jo, 2023) such as ChatGPT, with its new capabilities to generalise, summarise and provide human-like dialogue across almost every discipline, is set to disrupt the education sector from K-12 through to lifelong learning by challenging traditional systems and pedagogical approaches (Akgun & Greenhow, 2022; Aoun, 2017).

Artificial intelligence can be defined as the simulation of human intelligence and its processes by machines, especially computer systems, which encompasses learning (the acquisition of information and rules for using information), reasoning (using rules to reach approximate or definite conclusions), and flexible adaptation (Davis & Marcus, 2015; Kaplan & Haenlein, 2019; Russell & Norvig, 2010). In education, AI, or AIED, aims to "*make computationally precise and explicit forms of educational, psychological and social knowledge which are often left implicit*" (Self, 1998, p. 350). Therefore, the promise of AI to revolutionise education is predicated on its ability to provide adaptive and personalised learning experiences, thereby recognising and nurturing the unique cognitive capabilities of each student (Bulger, 2016). Furthermore, integrating AI into pedagogical approaches and practice presents unparalleled opportunities for efficiency, global reach, and the potential for the democratisation of education unattainable by traditional approaches.

AIED encompasses a broad spectrum of applications, from adaptive learning platforms that curate customised content to fit individual learning styles and paces (Kulik, 2003) to AI-driven analytics tools that forecast student performance and provide educators with actionable insights (Picciano, 2012). Moreover, recent developments in AIED have expanded the educational toolkit to include chatbots for student support, natural language processing for language learning, and machine learning for automating administrative tasks, allowing educators to focus more or exclusively on teaching and mentoring (Winkler & Söllner, 2018). These tools have recently converged into multipurpose, generative pre-trained transformers (GPTs). These GPTs are large language models (LLMs) utilising transformers to combine large language data sets and immense computing power to create an intelligent model which, after training, can generate complex, advanced, human-level output (Vaswani et al., 2017) in the form of text, images voice and video. These models are capable of multi-round human-computer dialogues, continuously responding with novel output each time users input a new prompt due to having been trained with data from the available corpus of human knowledge, ranging from the physical and natural sciences through medicine to psychiatry.

This convergence highlights that a step change has occurred in the capabilities of AI to act not only as a facilitator of educational content but also as a dynamic tool with *agentic* properties capable of interacting with stakeholders at all levels of the educational ecosystem, enhancing and potentially disrupting the traditional pedagogical process. Recently, the majority of the conversation within the current literature concerning AIED is focused on the aspect of cheating or plagiarism (Abd-Elaal et al., 2022; Sharples, 2022; Yeo, 2023), with some calls to examine the ethics of AI (Borenstein & Howard, 2021). This focus falls short of addressing the multidimensional, multi-stakeholder nature of AI-related issues in education. It fails to consider that AI *is already here*, accessible, and proliferating. It is this accessibility and proliferation that motivates the research presented in this manuscript. The release of generative AI globally and its application within education raises significant ethical concerns regarding data privacy, AI agency, transparency, explainability, and additional psychosocial factors such as confidence and trust, as well as the acceptance and equitable deployment of the technology in the classroom (Xia et al., 2022).

We, therefore, seek to understand the level of acceptability of AI within education for all stakeholders: students, teachers, parents, school staff and principals. Using factors derived from the explainable AI literature (Kelly et al., 2023) and the UNESCO framework for AI in education (Chan, 2023), we present research that investigates the role of agency, privacy, explainability and transparency in shaping the perceptions of global utility, individual usefulness, confidence, justice and risk toward AI and the eventual acceptance of AI in the classroom.

The remainder of this manuscript is organised as follows: we provide an overview of the constructs under investigation as they relate to AI within education, detail the study methodology, report results, and discuss them in context and more broadly.

# 2 Background

## 2.1 Agency: Autonomy and Control

In an educational setting, the deployment of AI has the potential to redistribute agency over decision-making between human actors (teachers and students) and algorithmic systems or autonomous agents. As AI systems come to assume roles traditionally reserved for educators, the negotiation of autonomy between educator, student and this new third party becomes a complex balancing act in many situations, such as personalising learning pathways, curating content, and even evaluating student performance (Darvishi et al., 2024; Holmes et al., 2023).

Educational professionals face a paradigm shift where the agency afforded to AI systems must be weighed against preserving the educators' pedagogical authority and expertise (Selwyn, 2019). However, this is predicated on human educators providing additional needs such as guidance, motivation, facilitation, and emotional investment, which may not hold as AI technology develops (Schiff, 2021). That is not to say that AI will supplant the educator in the short term, but rather, it highlights the requirement for carefully calibrating AI's role within the pedagogical process.

Student agency, defined as the individual's ability to act independently and make free choices (Bandura, 2001), can be compromised or enhanced by AI. While AI can personalise learning experiences, adaptively responding to student needs, thus promoting agency (Xie et al., 2019), it can conversely reduce student agency through over-reliance, whereby AI-generated information may diminish students' critical thinking and undermine the motivation toward self-regulated learning, leading to a dependency (Tsai et al., 2019).

Moreover, in educational settings, the degree of agency afforded to AI systems, i.e., its autonomy and decision-making capability, raises significant ethical considerations at all stakeholder levels. A high degree of AI agency risks producing "automation complacency" (Parasuraman & Manzey, 2010), where stakeholders within the education ecosystem, from parents to teachers, uncritically accept AI guidance due to overestimating its capabilities. Whereas a low degree of agency essentially hamstrings the capabilities of AI and the reason for its application in education. Therefore, ensuring that AI systems are designed and implemented to support and enhance human agency through human-centred alignment and design rather than replacing it requires thorough attention to the design and deployment of these technologies (Buckingham Shum et al., 2019).

In conclusion, educational institutions must navigate the complex dynamics of assigned agency when integrating AI into pedagogical frameworks; this will require careful consideration of the balance between AI autonomy and human control to prevent the erosion of the agency of stakeholders at all levels of the education ecosystem and, thus, increase confidence and trust in AI as a tool for education.

## 2.2 Confidence: Building Trust in AI

Establishing confidence in AI systems is multifaceted, encompassing the ethical aspects of the system, the reliability of AI performance, the validity of its assessments, and the robustness of data-driven decision-making processes (Kroll, 2015; Rader et al., 2018). Thus, confidence in AI systems within educational contexts centres on the capacity of AI systems to operate reliably and contribute meaningfully to educational outcomes.

Building confidence in AI systems is directly linked to the consistency of their performance across diverse pedagogical scenarios (Romero & Ventura, 2020). Consistency and reliability are judged by the AI system's ability to function without frequent errors and sustain its performance over time (Rauber et al., 2023). Thus, inconsistencies in AI performance, such as system downtime or erratic behaviour, may alter perceptions of utility and significantly decrease user confidence.

AI systems are increasingly employed to grade assignments and provide feedback, which are activities historically under the supervision of educators. Confidence in these systems hinges on their ability to deliver feedback that is precise, accurate and contextually appropriate (O'neil, 2017). The danger of misjudgement by AI, particularly in subjective assessment areas, can compromise its credibility (Balfour, 2013), increasing risk perceptions for stakeholders such as parents and teachers and directly affecting learners' perceptions of how fair and just AI systems are.

AI systems and the foundation models they are built upon are trained over immense curated datasets to drive their capabilities (Bommasani et al., 2021). The provenance of these data, the views of those who curate the subsequent training data and how that data is then used within the model (that creates the AI) is of critical importance to ensure bias does not emerge when the model is applied (Borenstein & Howard, 2021; Holmes et al., 2021). To build trust in AI, stakeholders at all levels must have confidence in the integrity of the data used to create an AI, the correctness of analyses performed and any decisions proposed or taken (Dietvorst et al., 2015). Moreover, the confidence-trust relationship in AI-driven decisions requires transparency about data sources, collection methods, and explainable analytical algorithms (Ribeiro et al., 2016).

Therefore, to increase and maintain stakeholder confidence and build trust in AIED, these systems must exhibit reliability, assessment accuracy, and transparent and explainable decision-making. Ensuring these attributes requires robust design, testing, and ongoing monitoring of AI systems, the models they are built upon, and the data used to train them.

## 2.3 Trust in AI: An Essential Factor Toward Acceptance

Trust in AI is essential to its acceptance and utilisation at all stakeholder levels within education. Confidence and trust are inextricably linked (Karran et al., 2022), representing a feedback loop wherein confidence builds towards trust and trust instils confidence, and the reverse holds that a lack of confidence fails to build trust. Thus, a loss of trust decreases confidence. Trust in AI is engendered by many factors, including but not limited to the transparency of AI processes, the alignment of AI functions with educational ethics, including risk and justice, the explainability of AI decision-making, privacy and the protection of student data, and evidence of AI's effectiveness in improving learning outcomes (Kashive et al., 2020; Kroll, 2015; Lee et al., 2021).

Standing as a proxy for AI, studies of trust toward automation (Hoff & Bashir, 2015; Lee & See, 2004) have identified three main factors that influence trust: performance (how automation performs), process (how it accomplishes its objective), and purpose (why the automation was built originally). Accordingly, educators and students are more likely to trust AI if they can comprehend its decision-making processes and the rationale behind its recommendations or assessments (Burrell, 2016). Thus, if AI operates opaquely as a "black box", it can be difficult to accept its recommendations, leading to concerns about its ethical alignment. Therefore, the dynamics of stakeholder trust in AI hinges on the assurance that the technology operates transparently and without bias, respects student diversity, and functions fairly and justly (Dignum, 2018).

Furthermore, privacy and security directly feed into the trust dynamic in that educational establishments are responsible for the data that AI stores and utilises to form its judgments. Tools for AIED are designed, in large part, to operate at scale, and a key component of scale is cloud computing, which involves resource sharing, which refers to the technology and the data stored on it (Amo Filvá et al., 2021). This resource sharing makes the boundary between personal and common data porous, which is viewed as a resource that many technology companies can use to train new AI models or as a product (Zuboff, 2023). Thus, while data breaches may erode trust in AIED in an immediate sense, far worse is the hidden assumption that all data is common. However, this issue can be addressed by stakeholders at various levels through ethical alignment negotiations, robust data privacy measures, security protocols, and policy support to enforce them (Chan, 2023; Nguyen et al., 2023).

Accountability is another important element of the AI trust dynamic and one inextricably linked to agency and the problem of control. It refers to the mechanisms in place to hold system developers, the institutions that deploy AI, and those that use AI responsible for the functioning and outcomes of AI systems (Kroll, 2015). The issue of who is responsible for the decisions or mistakes AI makes is an open question heavily dependent on deep ethical analysis. However, it is of critical and immediate importance, particularly in education, where the stakes include the quality of teaching and learning, the fairness of assessments, and the well-being of students.

In conclusion, trust in AI is an umbrella construct that relies on many factors interwoven with ethical concerns. The interdependent relationship between confidence and trust suggests that the growth of one promotes the enhancement of the other. At the same time, their decline, through errors in performance, process, or purpose, leads to mutual erosion. The interplay between confidence and trust points towards explainability and transparency as potential moderating factors in the trust equation.

## 2.4 Transparency and Explainability: Methods to Build Trust

The contribution of explainability and transparency towards trust in AI systems is significant, particularly within the education sector; they enable stakeholders to understand and rationalise the mechanisms that drive AI decisions (Guidotti et al., 2018). Comprehensibility is essential for educators and students to not only follow but also critically assess and accept the judgments made by AI systems (Arrieta et al., 2020; Gunning et al., 2019). Transparency gives users visibility of AI processes, which opens AI actions to scrutiny and validation (Lipton, 2018).

Calibrating the right balance between explainability and transparency in AI systems is crucial in education, where the rationale behind decisions, such as student assessments and learning path recommendations, must be clear to ensure fairness and accountability (Khosravi et al., 2022; Rader et al., 2018). The technology is perceived to be more trustworthy when AI systems articulate, in an accessible manner, their reasoning for decisions and the underlying data from which they are made (Rosenfeld & Richardson, 2019). Furthermore, transparency allows educators to align AI-driven interventions with pedagogical objectives, fostering an environment where AI acts as a supportive tool rather than an inscrutable authority (Bearman & Ajjawi, 2023; Leslie, 2019; Niemi, 2021).

Moreover, the explainability and transparency of AI algorithms are not simply a technical requirement but also a legal and ethical one, depending on interpretation, particularly in light of regulations such as the General Data Protection Regulation (GDPR), which posits a "right to explanation" for decisions made by automated systems (Felzmann et al., 2019; Goodman & Flaxman, 2017; Hamon et al., 2022). Thus, educational institutions are obligated to deploy AI systems that not only perform tasks effectively but also provide insights into their decision-making processes in a transparent manner (Sovrano et al., 2020; Wachter et al., 2017).

In sum, explainability and transparency are critical co-factors in the trust dynamic, where trust appears to be the most significant factor toward the acceptance and effective use of AI in education. Systems that employ these methods enable stakeholders to understand, interrogate, and trust AI technologies, ensuring their responsible and ethical use in educational contexts.

## 2.5 Toward AI Acceptance in Education

When taken together, this discussion points to the acceptance of AI in education as a multifaceted construct, hinging on a harmonious yet precarious balance of agency, confidence, and trust underpinned by the twin pillars of explainability and transparency. Agency involving the balance of autonomy between AI, educators, and students requires careful calibration between AI autonomy and educator control to preserve pedagogical integrity and student agency, which is vital for independent decision-making and critical thinking. Accountability, closely tied to agency, strengthens trust by ensuring that AI systems are answerable for their decisions and outcomes, reducing risk perceptions. Trust in AI and its co-factor confidence are fundamental prerequisites for AI acceptance in educational environments. The foundation of this trust is built upon factors such as AI's performance, the clarity of its processes, its alignment with educational ethics, and the security and privacy of data. Explainability and transparency are critical in strengthening the trust dynamic. They provide stakeholders with insights into AI decision-making processes, enabling understanding and critical assessment of AI-generated outcomes and helping to improve perceptions of how just and fair these systems are.

However, is trust a one-size-fits-all solution to the acceptance of AI within education or is it more nuanced where different AI applications require different levels of each factor on a case-by-case basis and for different stakeholders? This research seeks to determine to what extent each factor contributes to the acceptance and intention to use AI in education across four use cases from a multi-stakeholder perspective.

## 3 Materials and Methods

For the purpose of this study, which included both a manipulation check and a broader multi-stakeholder investigation, we utilised an experimental vignette methodology (Aguinis & Bradley, 2014). The experimental vignette methodology (EVM) consisted of presenting participants with four carefully developed scenarios anchored in current and future educational reality to assess the effect of four independent variables Agency, Transparency, Explainability and Privacy upon the perceptions of Global Utility (GU), Individual Usefulness (IU), Justice, Confidence, Risk, and the Intention to Use (ITU).

### 3.1 Vignette Development

Before settling on the specific formulation and wording for the scenarios and subsequent survey (the heart of the data collection), we conducted a series of small focus groups utilising a "foresight" methodology proposed by (Inayatullah, 2008). The purpose was to gain insight related to the current perceptions of AI by students, parents, high school teachers and school directors. These insights were meant to qualify perceptions of AI in the classroom, in the education literature (Chounta et al., 2022; Kashive et al., 2020) and to inform the research team by providing current insights related to the current lived experience of five sixteen-year-old students (3 male), four teaching staff (3 female), two school directors (2 Female) and three parents (3 Male) of students in high school.

The focus groups were structured around four scenarios ranging from probable to futuristic: an AI that helps with grading, an AI that personalises educational content to student needs, an AI that monitors student learning, and finally, an AI that supports and helps boost student performance. To finish every session, we asked a final question: "Where do you set the hard boundary on integrating AI into the classroom?"

Following each focus group, the wording of the scenarios in development was iterated until the final set of four scenarios and four independent variables was approved (see Table 2), resulting in a 2x2x2x2 design the complete set of vignettes is available in Appendix 1. The focus groups and subsequent surveys were approved by the author's institution's research ethics board.

## 3.2 Survey Development

Based on literature survey and focus group insights, a survey was developed to assess the impact of agency, transparency, explainability and privacy upon the perceptions of global utility adapted from the utilitarian perspective as "that which provides the greatest good for the greatest number" (Sinnott-Armstrong, 2023); perceived individual usefulness as the degree to which a person believes that using a particular system or technology will enhance their job performance or productivity (Davis, 1989); justice taken from the personal relativist standpoint (Baghramian & Coliva, 2019) as how "just" an AI is perceived in its judgements; confidence as the belief that someone or something is honest, reliable, good, and effective, or the desire to depend on someone or something for security (Safa & Von Solms, 2016; Ye et al., 2019); risk as the combination of uncertainty and the seriousness of an outcome in relation to performance, safety, and psychological or social uncertainties (Ye et al., 2019), and intention to use as the likelihood or willingness of individuals to use a particular information system or technology (Davis, 1989).

*Table 1. Dependent variables and corresponding scales*

| Dependent Variable | Scale | # Items | Adapted from |
| --- | --- | --- | --- |
| Perceived global utility | Likert 1-5 | Four items | (Hyman, 1996; Reidenbach & Robin, 1988, 1990) |
| Perceived individual usefulness | Likert 1-5 | Seven items | (Ye et al., 2019) |
| Justice | Likert 1-5 | Three Items | (Hyman, 1996; Reidenbach & Robin, 1988, 1990) |
| Confidence (Trust) | Likert 1-5 | Three items | (Ye et al., 2019) |
| Risk | Likert 1-5 | Four items | (Ye et al., 2019) |
| Intention to Use | Likert 1-5 | Five items | (Ye et al., 2019) |

The survey instrument was created using a combination of verified scales adapted from the ethics and information systems literature (see Appendix 2). Table 1 shows the overall number of items and provenance of the original scales utilised for adaptation. All scales were translated from English to French, using a 3-way majority vote with an IRR of 0.82 and modified for the target stakeholder.

*Table 2. Two vignette examples: Correct-AI and Answer-AI scenarios presented to participants in this study.*

| Scenario - Context 1 – *Correct-AI* | Scenario - Context 2 – *Answer-AI* |
|---|---|
| Imagine a high school with *Correct-AI*, an artificial intelligence (AI) technology used to mark students' exams. *Correct-AI* mobilises artificial intelligence algorithms capable of analysing complex written answers, understanding graphs and equations, and evaluating problem-solving strategies. *Correct-AI* is then able to automatically assign grades and provide feedback to students. Here is more relevant information about *Correct-AI*: Grades provided by *Correct-AI* [**Agency-High**: are not validated; **Agency-Low**: are validated] by a teacher. *Correct-AI* [**Privacy-Low**: uses learner personal data beyond academic performance (e.g., gender, native language, diagnosis of learning difficulties, etc.); **Privacy-High**: does not use learner personal data beyond academic performance] in the correction algorithm. *Correct-AI* Provides the [**Teacher**, **Student**, **Parent**] with [**Explainability-Low**: only the note at the end of the correction without further explanation; **Explainability-High**: the note and a full explanation of the correction process]. It [**Transparency-Low**: is not communicated; **Transparency-High:** is clearly communicated] [to the **Teacher**, **Student**, **Parent**] what data is collected and how it is used by *Correct-AI*. | Imagine a high school using *Answer-AI*, a chatbot (i.e., a conversational agent based on artificial intelligence) to answer students' questions in various courses outside class hours. *Answer-AI* is based on state-of-the-art artificial intelligence algorithms that can automatically maintain a text conversation with students without the help of a human. This chatbot is able to understand the meaning and context of the conversation, and is therefore able to provide students with sensible, course-appropriate answers. Here is more relevant information about Answer-AI: Answer-AI formulates its answers to students' questions [**Agency-High**: without being supervised; **Agency-Low**: without being supervised] by the teacher. Answer-AI [**Privacy-Low**: shares the content of the conversation with the teacher and principals; **Privacy-High**: does not share the content of the conversation with anyone else]. Answer-IA provides the student with [**Explainability-Low**: only the answer to his/her question, without further explanation; **Explainability-High**: the answer to his/her question and the explanation for arriving at this answer]. The student [**Transparency-Low**: is not informed; **Transparency-High:** is informed] that he *or* she is communicating with a chatbot and not a human. |

## 3.3 Manipulation Check

In the second phase of the study carried out in 2022, the four scenarios were tested for manipulation validity. The scenarios were administered using Qualtrics and distributed over the Prolific (Prolific Inc., UK) online platform four times, each time a different single context was presented with a different combination of independent variable directionality, i.e., low *agency*, low *privacy*, high *explainability*, high *transparency* to capture views from 500 participants across North America. Target participants were adults older than 18, with at least one child currently enrolled in compulsory secondary education aged between 14 and 17 years. All participants provided digital consent and could cease completing the survey at any time. See Appendix 1 for the questions used for the manipulation check.

*Table 3. Participant distribution across scenarios*

| Scenario | N | Sex |
|---|---|---|
| **Scenario 1 - *Correct-AI*** | 62 | 36 Female |
| **Scenario 2 - *Answer-AI*** | 59 | 33 Female |
| **Scenario 3 - *Tutor-AI*** | 59 | 31 Male |
| **Scenario 4 - *Emotion-AI*** | 66 | 35 Female |

In total, 246 participants provided valid data for analysis across the four scenarios (see *Table 3*), the manipulation check data was analysed using the Wilcoxon sum rank test, and the results indicated a strong effect of the manipulations (see *Figure 1*). That is, there were significant differences between low and high agency (z=9.08, p<.0001), privacy (z=8.26, p<.0001), explainability (z=8.04, p<.0001) and transparency (z=-7.11, p<.0001), allowing the next stage of the data collection to proceed.

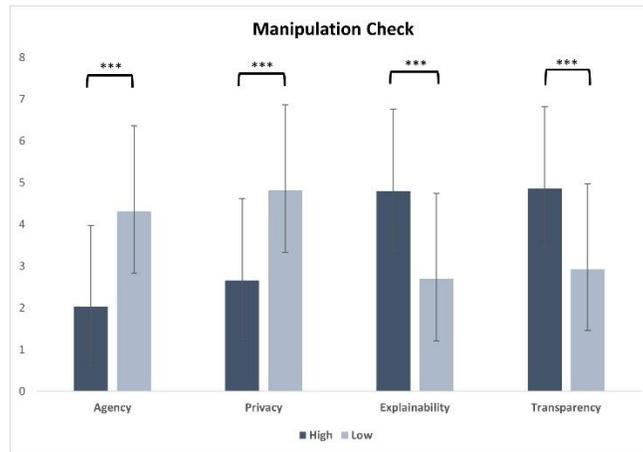

*Figure 1. Results from the manipulation check \*=p<0.05, \*\*=p<0.005, \*\*\*=p<0.0001*

## 3.4 Data Collection

The final data collection phase was partially distributed by AlloProf (AlloProf Inc), a non-profit education sector organisation providing academic support services at all levels of the education ecosystem in Quebec, Canada. In addition, we also utilised a social media campaign composed of advertisements in schools, Facebook private groups (parents, teachers, and school directors) and LinkedIn to recruit a multi-stakeholder pool of participants from within Quebec, Canada only. The survey was presented in both English and French. Individual participants were exposed to one of the four scenarios and the complete survey. In total, 4073 participants participated in the survey (see Table 4). The survey was embedded with several attention checks to ensure quality participation. Data for this part of the survey exercise were only included if the survey was completed in full and all attention checks were passed. After removal of failed attention checks and incomplete survey data, the sample consisted of 608 students aged 16-18 (µ= 17.4, 357 Female, 190 Male, 61 Other), 381 parents (306 Female, 58 Male, 17 Other) with at least one child in mandatory education, 203 teachers (138 Female, 59, Male, 6 Other) currently working in primary or secondary education and 20 directors (12 Female, 8 Male) of secondary education institutions. A prize draw of 1 iPad was also offered to encourage student participation. However, due to the low sample size, the data for school directors was omitted from the mediation analysis.

*Table 4. Participant distribution and total respondents*

|  | *Student* | *parent* | *Teacher* | *Director* | *Total Respondents* |
|---|---|---|---|---|---|
| *No. of responses* | 2635 | 900 | 486 | 52 | 4073 |
| *No. of finished responses* | 1454 | 433 | 238 | 23 | 2148 |
| *No. of valid responses* | 608 | 381 | 203 | 20 | 1212 |

As previously stated, each participant was exposed to only one scenario, and the complete survey instrument, shown in Table 5, is the distribution of participants across the four scenarios.

Table 5. Distribution of respondents across the four scenarios

| Role | Scenario1 | Scenario 2 | Scenario 3 | Scenario 4 | Total |
|---|---|---|---|---|---|
| Director | 4 | 7 | 4 | 5 | 20 |
| Parent | 110 | 89 | 82 | 100 | 381 |
| Teacher | 51 | 50 | 52 | 50 | 203 |
| Student | 184 | 135 | 146 | 143 | 608 |
| Total | 349 | 281 | 284 | 298 | 1212 |

The survey instrument used in this study consisted of six subscales (see Table 1); after analysis, Cronbach's alpha showed high internal consistency for all the scales utilised. The global utility subscale consisted of 5 items ($\alpha$= .81), the perceived usefulness consisted of 7 items ($\alpha$= .79), the justice subscale consisted of 3 items ($\alpha$= .85), the confidence subscale consisted of 3 items ($\alpha$= .71), the risk subscale consisted of 4 items ($\alpha$= .74), and the intention to use subscale consisted of 5 items ($\alpha$= .90).

## 4 Analysis

A multiple regression analysis was performed as a first step to examine the direct effects of the independent variables' explainability, agency, privacy, and transparency on the dependent variable intention to use. The model significantly predicted intention to use, $F(4, 1207)=4.683$, $p<.001$), and accounted for approximately 1.5% of the variance in intention to use ($R^2$=.015, Adjusted $R^2$=.012). Agency was shown to be a significant predictor, with a higher level of Agency associated with a lower intention to use ($\beta$=-.2519, p<.001). However, Transparency ($\beta$ = .0168, $p$=.799), Explainability ($\beta$=.1189, $p$=.071), and Privacy ($\beta$=.0718, $p$=.275) were not significant predictors of intention to use, indicating that other factors, may mediate the intention to use AI in the classroom.

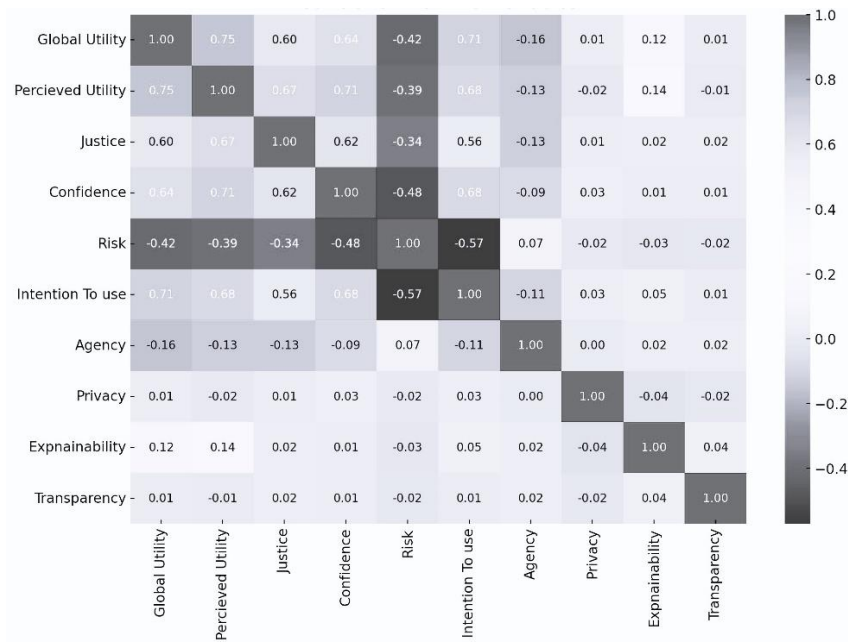

Figure 2. Correlation matrix of all variables under investigation

To determine the next steps in the data analysis, we performed a correlation analysis of all variables (see Figure 2), which reported moderate correlations between dependent variables. For instance,

'perceived utility' and 'global utility' show a positive correlation, suggesting that as perceptions of (individual) utility increase, perceptions of overall utility also tend to be higher. However, no strong correlations were reported between the independent variables (agency, privacy, explainability, transparency) and the dependent variables, indicating that these factors might indirectly influence the dependent variables.

We conducted a mediation analysis (see Figure 3) to determine these influences through an examination of the direct and indirect effects of the independent variables *explainability*, *agency*, *privacy,* and *transparency* on the dependent variable *intention to use* through *global utility*, *individual utility*, *justice*, *confidence*, and *risk* from a multi-stakeholder perspective using Hayes (Hayes, 2017) Process Macro (Model 4) in SPSS. For all analyses performed, the confidence level for all confidence intervals in the output is 95.0%, and the number of bootstrap samples for percentile bootstrap confidence intervals was set to 5000. This analysis models the direct effect of C' of each IV $X$ on DV $Y$ and the indirect effect of $X_i$ on $Y$ through mediator(s) $M_i$ ($a_i*b_i$); the PROCESS macro allows up to ten mediators to operate in parallel.

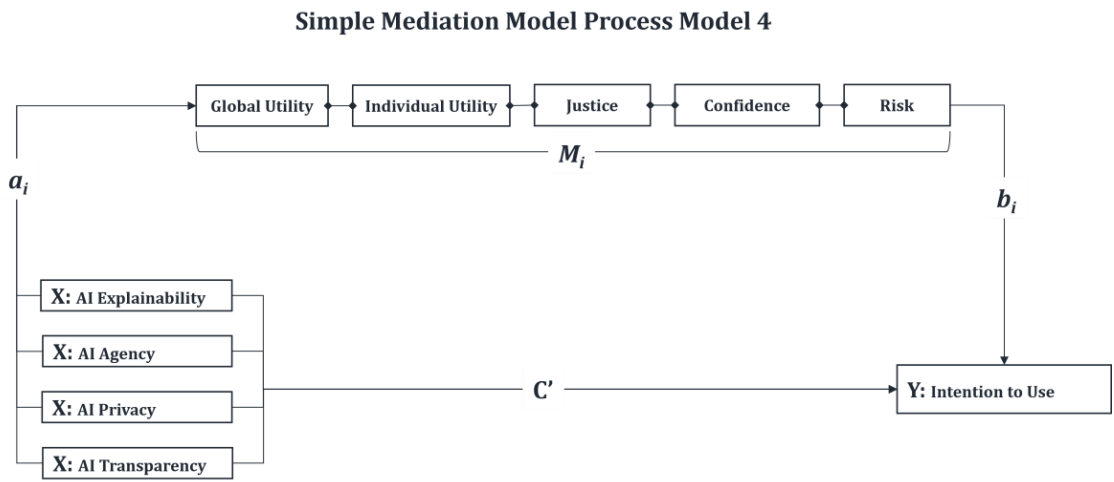

*Figure 3. Simple mediation model: direct effect C' of $X_i$: on Y: and indirect effects of $X_i$: mediated by $M_i$: on Y:*

## 5 Results

This section is structured to provide a detailed report of the results reported by the simple mediation model analysis by stakeholder group and scenario.

### 5.1 Mediators of intention to use: Students.

#### 5.1.1 Scenario one: Correct AI

The analysis performed on student data exposed to scenario one, "Correct-AI", reported no direct effect of any independent variable upon the dependent variable intention to use. However, some mediation effects were reported when considering the effect of agency through perceptions of GU and justice upon ITU.

The analysis revealed that the level of agency significantly negatively impacted GU ($\beta$ = -.4261, p = .0011), indicating that higher levels of agency were associated with lower levels of perceived GU. Furthermore, GU was found to have a significant positive impact on ITU ($\beta$ = .3402, p = .0001), suggesting that higher levels of both personal and overall benefit may lead to greater ITU.

Similarly, agency significantly negatively impacted justice (β = -.3768, p = .0261), indicating that higher levels of agency were associated with a lower perception of how "just" the system would be. Justice was found to significantly impact ITU (β = .1760, p = .0041), suggesting that students were concerned that AI with high agency would be less just. However, the direct effect of agency on ITU was not statistically significant (β = -.1056, p = .2982), indicating that the level of agency alone did not significantly predict intention to use.

An examination of the indirect effects of agency on ITU through GU and justice showed that the indirect effect through GU was statistically significant (Effect = -.1449, BootSE = .0606, BootLLCI = -.2805, BootULCI = -.0441), as was the indirect effect through justice (Effect = -.0576, BootSE = .0343, BootLLCI = -.1382, BootULCI = -.0047).

### 5.1.2 Scenario two: Answer-AI

The analysis performed on the data for students exposed to scenario two, "Answer-AI", reported no direct effect of any independent variable upon the dependent variable intention to use. However, a mediation effect was reported when considering the effect of privacy through the perception of GU upon the ITU.

The analysis reported that privacy had a significant positive impact on GU (β = .2987, p = .0439), indicating that higher levels of privacy were associated with higher levels of both personal and overall benefit. GU was found to have a significant positive impact on ITU (β = .3159, p = .0007), indicating that higher levels of both personal and overall benefit may lead to greater ITU.

However, the direct effect of privacy on intention to use was not statistically significant (β = .0369, p = .7346), indicating that privacy alone did not significantly predict ITU. In contrast, the indirect effect of privacy on intention through GU was found to be statistically significant (Effect = .0944, BootSE = .0586, BootLLCI = .0022, BootULCI = .2283).

### 5.1.3 Scenario three: Tutor-AI

The mediation analysis reported no direct or indirect effects for this scenario context.

### 5.1.4 Scenario four: Emotion-AI

The analysis performed on the data for students exposed to scenario four, "Emotion -AI", reported no direct effect of any independent variable upon the dependent variable ITU. However, a mediation effect was reported when considering the effect of explainability through the perception of GU upon ITU. Whereby explainability had a significant positive impact on GU (β = .2787, p = .0259), indicating that higher levels of explainability are associated with higher levels of personal and overall benefit. Moreover, GU was found to have a significant positive impact on ITU (β = .3089, p = .0039).

In this case the direct effect of explainability on ITU was not statistically significant (β = -.0429, p = .7187), highlighting that explainability alone did not significantly predict ITU. However, the indirect effect was found to be statistically significant (Effect = .0861, BootSE = .0513, BootLLCI = .0039, BootULCI = .2018).

## 5.2 Mediators of intention to use: Parents.

The mediation analysis performed for parents exposed to all four scenarios reported significant results only for scenario one (*Correct-AI* see Table 2) concerning the relationship between agency, the mediator, confidence, and ITU.

The results revealed that agency negatively impacted confidence (β = -.4668, p = .0236), indicating that higher levels of agency were associated with lower confidence levels in the context of an AI utilised to correct student work. Furthermore, confidence was found to have a significant positive impact on ITU (β = .5200, p < .001), indicating that greater confidence was associated with higher

ITU. However, the direct effect of agency on ITU was not statistically significant (β = .0708, p = .5968), suggesting that agency alone did not significantly predict ITU.

Further analysis revealed an indirect effect of agency on ITU through confidence. Specifically, the indirect effect of agency on ITU through confidence was found to be statistically significant (Effect = -.2428, BootSE = .1184, BootLLCI = -.4956, BootULCI = -.0347), suggesting that agency's influence on ITU is partially mediated by confidence.

## 5.3  Mediators of intention to use: Teachers.

Data analysis for the teacher stakeholder reported several significant results across multiple scenarios.

### 5.3.1  Scenario one: Correct-AI

For scenario one, the mediation analysis reported that explainability had a significant positive impact on global utility (β = .7161, p = .0057), indicating that higher levels of explainability were associated with higher levels of perceived personal and overall benefit. Additionally, GU was found to have a significant positive impact on ITU (β = .3639, p = .0286).

Additionally, explainability had a significant positive impact on perceived usefulness (β = .6634, p = .0059), indicating that greater levels of explainability were associated with higher perceptions of personal usefulness. Moreover, perceived usefulness significantly impacted ITU (β = .6205, p = .0087).

However, the direct effect of explainability on ITU was not statistically significant (β = -.0215, p = .9366), indicating that explainability alone did not significantly predict ITU. In contrast, the mediation analysis reported that both the indirect effect through GU and perceived usefulness were found to be statistically significant (Effect = .2606, BootSE = .1717, BootLLCI = .0215, BootULCI = .6790) and (Effect = .4116, BootSE = .2456, BootLLCI = .0370, BootULCI = .9902) respectively. This result suggests that the influence of explainability on ITU is partially mediated by GU and perceived usefulness.

### 5.3.2  Scenario two: Answer-AI

For scenario two, the mediation reported that both agency and explainability had a significant impact on perceived GU negatively (β = -.5969, p = .0271) in the case of agency and positively (β = .8700, p = .0008) in the case of explainability.

Furthermore, in the case of agency, GU was found to have a significant positive impact on intention to use (β = .5590, p = .0040). However, the direct effect of agency was not statistically significant (β = -.2283, p = .2847), indicating that agency alone did not significantly predict intention to use. Whereas, the indirect effect of agency on ITU through GU was statistically significant (Effect = -.3337, BootSE = .2091, BootLLCI = -.8033, BootULCI = -.0050).

Additionally, in the case of explainability, GU was found to have a significant positive impact on intention to use (β = .5970, p = .0030). Similarly, the direct effect of explainability on intention to use was not statistically significant (β = -.1479, p = .5028), indicating that explainability alone did not significantly predict intention to use. At the same time, the indirect effect of explainability on intention to use through GU was found to be statistically significant (Effect = .5194, BootSE = .2455, BootLLCI = .1120, BootULCI = 1.0618).

The mediation analyses for this scenario indicate that GU plays a mediating role for agency, explainability, and the intention to use AI in this context. While agency and explainability did not directly predict ITU, they influenced ITU indirectly through their effects on GU. These findings suggest the importance of held or considered utilitarian values as a mediator in understanding the impact of agency and explainability on teachers' intention to use AI in this context.

### 5.3.3 Scenario three: Tutor-AI
The mediation analysis reported no direct or indirect effects for this context.

### 5.3.4 Scenario four: Emotion-AI
For scenario four, the mediation analysis reported that privacy had a significant negative impact on confidence ($\beta = -.6581$, $p = .0012$), indicating that higher levels of privacy were associated with lower confidence levels in the context of an AI that supports emotional health. Additionally, confidence significantly impacted ITU ($\beta = .4744$, $p = .0447$). However, the direct effect of privacy on ITU was not statistically significant ($\beta = .1118$, $p = .6144$), indicating that privacy alone did not significantly predict ITU or accept the use of AI for this purpose. Whereas, the indirect effect of privacy on intention through confidence was statistically significant (Effect = -.3122, BootSE = .1528, BootLLCI = -.6559, BootULCI = -.0511).

The results of this mediation analysis indicate that, for teachers, the level of privacy significantly influences confidence, which has a significant positive impact on ITU, where privacy is low. While the level of privacy did not directly predict ITU, it indirectly affects ITU or acceptance of AI for this purpose through its impact on confidence.

## 6 Discussion
When taken as a whole, the results indicate that different stakeholder groups have varying degrees of acceptance and trust across AI applications. The key mediators influencing the intention to use AI within education in the current study include global utility, justice, and confidence, which vary according to the level of AI agency, transparency, and explainability.

The results indicated that students' intention to use AI in education was influenced by their perceptions of AI's global utility and justice. For instance, in the Correct-AI scenario where AI is employed to correct student work, a higher level of agency in AI was associated with lower perceptions of global utility and justice. This suggests that students were concerned about AI systems that were too autonomous, perceiving them as less just and beneficial. The results indicate that privacy was valued in the second scenario, Answer-AI, a conversational agent that answers students' questions in various courses outside class hours. However, privacy did not directly influence student/learner intention to use but rather was positively associated with higher perceived global utility. This finding indicates that an AI that keeps student queries private was perceived to be beneficial. For the fourth scenario, Emotion-AI, a conversational agent that provides emotional support to students, explainability positively impacted global utility, indicating that students perceived AI that was more explainable AI to be more beneficial, especially in the context of emotional support. However, explainability did not directly predict the intention to use AI, highlighting the importance of perceived benefit.

Moving to the second stakeholder group, parents, the results from the mediation analysis reported a significant relationship between agency and confidence, specifically in the Correct-AI scenario. Higher levels of AI agency led to lower confidence, negatively impacting their intention to use or allow AI for correcting student work. This finding indicates that parents might be more accepting of AI in education if they have greater confidence in the system's ability to act responsibly and fairly.

Perceptions about the other three scenarios were neutral, showing no significant variance, potentially due to the nature of the AI's in question, which work to the benefit of a student alone, where the Correct-AI would have a direct impact on educational outcomes should autonomous AI judgements prove negative, which in turn would have far-reaching effects upon the future employment or academic prospects of the child in question.

For the third stakeholder group, teachers, the mediation analysis showed that the intention to use was influenced by their perceptions of global utility and perceived usefulness, particularly in the context of explainability. Scenario one Correct-AI reported that higher explainability was associated with a greater perception of global utility and usefulness. For scenario two, Answer-AI, both agency and explainability influenced perceived global utility. While agency had a negative influence, explainability positively influenced this perception, indicating that teachers prefer AI systems that are less autonomous but more explainable in this instance. Moving to scenario four, Emotion-AI, it was reported that higher privacy levels had a negative impact on confidence, indicating that teachers were concerned about AI systems with high privacy levels handling the emotional needs of students. Whereas confidence positively influenced their intention to use AI, suggesting the need for a balance between privacy and trustworthiness.

These results reflect the complex and varied perceptions of AI among different stakeholder groups in the educational ecosystem. The findings suggest that while there is a general trend towards valuing transparency, explainability, and privacy in AI systems, there are, however, specific concerns and preferences that differ significantly across scenarios and stakeholder groups.

## 6.1  Key Mediators of AI Acceptance

Several key mediators of AI acceptance emerged from the current study: global utility, confidence, and justice. It appears that global utility was a significant factor influencing stakeholders' intentions to use AI across multiple scenarios, indicating that perceptions of the overall benefit of AI, from a multi-tiered perspective, may have cohered into an aggregative consequentialist assessment, i.e., the net "total" of personal benefits or harms coupled more broadly with benefits or harms to the educational ecosystem and by consequence society (Sen, 1982; Sinnott-Armstrong, 2023). Thus, on the strength of the results in this study alone, a positive balance of this utilitarian assessment is crucial for the acceptance of AI in educational settings and, perhaps, the acceptance of AI more broadly. However, this generalisation is a matter of public discourse and lies far outside the scope of this article.

Confidence in AI systems also emerged as a significant mediator, especially for parents in the case of Correct-AI and teachers in the case of Emotion-AI. The level of confidence, influenced by factors such as level of agency and privacy, impacted their willingness to use or accept AI in educational contexts. This result underscores the importance of building reliable and trustworthy AI systems. In the case of confidence, the agency level assigned to an AI application is entwined with perceptions of how much data it will share externally and how that data will be used. From a teaching and administrative perspective, such data is valuable for predictive learning analytics to identify pedagogical issues and related needs, ostensibly so that learning content can be adapted within the boundaries of those needs (Kizilcec, 2023).

The results further highlighted that in some scenarios, the level of transparency and agency in AI systems was a significant concern. For instance, parents' low transparency was associated with higher perceived risk and, in the case of students' injustice. In comparison, AI, with greater transparency, reduced these perceptions. Additionally, the level of agency of AI systems affected stakeholders' confidence and perceptions of justice, indicating a need for a balanced approach in AI design and deployment.

Seen through a lens consisting of AI transparency and agency mediated by confidence, justice, privacy and risk, a question of bias concerning fairness emerges (Madaio et al., 2022). Conceivably, AI can serve the needs of students without resorting to sharing private information, reducing the need for classification and labelling and potentially improving the learning experience for the individual. However, education, like any industry, is driven by the economic forces installed at all levels of civil society (Davies & Bansel, 2007), and those "forces" through the enforcement of informational framing

(Kuypers, 2009) dictate where resources are directed, how individuals are classified and what current and future value they will provide to the state and beyond. This creates tension between the "egalitarian ideal" distribution of educational resources and those allocated to the for-profit organisations that are increasingly integral to the education system (Lobera et al., 2020; Newfield, 2023). Thus, this tension potentially coalesces into bias; students who do not succeed are classified as bad assets by an AI that is forced to share this information within the education ecosystem in which it is embedded. Unless there are subsidised programs to provide specialised help, they become marginalised while resources are allocated elsewhere simultaneously (Miller et al., 2018). Within this value-driven system, AI has the potential to take the role of championing egalitarian values and providing non-biased education at a pace suitable for all students. In light of the results from the current study, several questions emerge: at what level are teachers placing utilitarian value? Where is the greatest good? The student, the school, or state and corporate level stakeholders looking for return on investment.

## 6.2 Trust and Acceptance

The findings of this study, particularly in scenarios involving Correct-AI and Emotion-AI, echo concerns related to the potential of AI to redistribute agency (Holmes et al., 2023). In both scenarios, higher levels of AI agency were associated with lower perceptions of justice and general utility among students and decreased confidence among parents and teachers, reflecting the need to carefully calibrate AI's role in education (Buckingham Shum et al., 2019).

Within the educational ecosystem, confidence and trust in AI are central to its acceptance (Romero & Ventura, 2020); AI systems must demonstrate consistency, reliability and accuracy (Rauber et al., 2023) and contribute meaningfully to educational outcomes (Balfour, 2013) and do so transparently (Burrell, 2016) and without bias (Dignum, 2018). What emerged from the findings reflected this: confidence in AI, influenced by factors such as agency and privacy, played a significant role in stakeholders' intention to use AI. Specifically, concerning parents, confidence was shaped by the level of AI agency in Correct-AI, aligning with this emphasis on the reliability and accuracy of AI systems. Furthermore, trust, influenced by factors like explainability and transparency, significantly affected stakeholders' acceptance of AI. In the case of Emotion-AI, the level of transparency impacted student confidence, which in turn influenced the intention to use AI. This underscores the complex dynamic of confidence and trust in that they are interdependent constructs reliant on many factors and interwoven with ethical concerns. Thus, educational institutions need clear policies to deploy AI that address the problem of agency, build confidence and trust, and insist on explainable AI systems.

Transparency and explainability contribute significantly towards confidence and trust building in AI systems in education and beyond; they open up AI to scrutiny and validation (Lipton, 2018) for stakeholders at various levels of the educational ecosystem to understand and rationalise the decisions taken and judgements made by AI (Arrieta et al., 2020; Gunning et al., 2019). In line with these assertions, the results showed that explainability significantly influenced perceptions of global utility and intention to use AI among students and teachers. Consequently, educators must be trained not only in how to use AI tools but also in understanding the implications of their use and, of equal importance, any resistance to using AI tools.

Overarchingly, the results of the current study demonstrate that stakeholder perceptions of AI technology are critical determinants of their acceptance and eventual use and that this acceptance and use is not uniform across AI applications. It thus becomes essential to understand how stakeholder perceptions can be positively influenced through best practice AI design and development, information framing and policy reinforcement. Furthermore, the objective functions of AI applications developed for education needs must be clearly defined in terms of their utilitarian value and deployed responsibly. Artificial intelligence, as an advanced intelligent tool with ever-increasing agentic properties, has the potential to deliver on the liberal egalitarian ideal of education

(Brighouse, 2002), not only in Western culture but worldwide. However, to deliver this potential requires the promotion of a new cultural imaginary that is not anti-science and technology, based on fear, and that clearly demonstrates the benefits of AI within education on an individual, organisational, and global level.

## 6.3 Limitations & Conclusion

We set out initially to capture as wide an array of perceptions toward multiple AI applications from a multi-stakeholder perspective as possible. While we succeeded in this goal, sample sizes could be increased for some specific stakeholder groups, such as teachers and school directors. Given the small sample size of the school director stakeholder group, these data were excluded from the mediation analysis, thus reducing the breadth of inference concerning perceptions towards AI in the education ecosystem. A study focused solely on this group may be fruitful in adding to the strength of the inferences made within this study. Similarly, increasing the sample size for teachers may provide a more granular appraisal of the mediators of acceptance of AI in education.

Another issue involves the vignette methodology; while this method is instrumental in gathering initial perceptions, providing concrete examples of AI applications after deployment would be beneficial. This can be accomplished in future work with larger sample sizes once AI and the issues surrounding the technology have breached the public consciousness. As AI technology appears to be improving exponentially, this research will become essential to the education sector and beyond.

This study showed that stakeholders' acceptance of AI in education varied across different scenarios. Each AI application (e.g., Correct-AI, Answer-AI, Tutor-AI, Emotion-AI) elicited different reactions regarding global utility, justice, and confidence, reflecting the nuanced nature of trust and acceptance towards AI. Our findings suggest that applying a one-size-fits-all approach to AI integration in education is infeasible. Instead, careful consideration of specific AI applications and their impact on perceptions of global utility, justice, and confidence is needed to enhance stakeholder acceptance and trust in AI education solutions.

# Appendix 1 Scenarios.

| Scenario - Context 1 – *Correct-AI* | Scenario - Context 2 – *Answer-AI* |
|---|---|
| Imagine a high school with *Correct-AI*, an artificial intelligence (AI) technology used to mark students' exams. *Correct-AI* mobilises artificial intelligence algorithms capable of analysing complex written answers, understanding graphs and equations, and evaluating problem-solving strategies. *Correct-AI* is then able to automatically assign grades and provide feedback to students.<br><br>Here is more relevant information about *Correct-AI*:<br><br>Grades provided by *Correct-AI* [**Agency-High**: are not validated; **Agency-Low**: are validated] by a teacher.<br><br>*Correct-AI* [**Privacy-Low**: uses learner personal data beyond academic performance (e.g., gender, native language, diagnosis of learning difficulties, etc.); **Privacy-High**: does not use learner personal data beyond academic performance] in the correction algorithm.<br><br>*Correct-AI* Provides the [**Teacher**, **Student**, **Parent**] with<br><br>[**Explainability-Low**: only the note at the end of the correction without further explanation; **Explainability-High**: the note and a full explanation of the correction process].<br><br>It [**Transparency-Low**: is not communicated; **Transparency-High:** is clearly communicated] [to the **Teacher**, **Student**, **Parent**] what data is collected and how it is used by *Correct-AI*. | Imagine a high school using *Answer-AI*, a chatbot (i.e., a conversational agent based on artificial intelligence) to answer students' questions in various courses outside class hours. *Answer-AI* is based on state-of-the-art artificial intelligence algorithms that can automatically maintain a text conversation with students without the help of a human. This chatbot is able to understand the meaning and context of the conversation, and is therefore able to provide students with sensible, course-appropriate answers.<br><br>Here is more relevant information about Answer-AI:<br><br>Answer-AI formulates its answers to students' questions [**Agency-High**: without being supervised; **Agency-Low**: without being supervised] by the teacher.<br><br>Answer-AI [**Privacy-Low**: shares the content of the conversation with the teacher and principals; **Privacy-High**: does not share the content of the conversation with anyone else].<br><br>Answer-IA provides the student with [**Explainability-Low**: only the answer to his/her question, without further explanation; **Explainability-High**: the answer to his/her question and the explanation for arriving at this answer].<br><br>The student [**Transparency-Low**: is not informed; **Transparency-High:** is informed] that he *or* she is communicating with a chatbot and not a human. |

| Scenario - Context 3 – *Tutor-AI* | Scenario - Context 3 – *Emotion-AI* |
|---|---|
| Imagine a high school providing students with *Tutor-AI*, a personalised tutoring application based on artificial intelligence. *Tutor-AI* adjusts to the student's level dynamically, i.e. as he or she uses it and learns. In fact, *Tutor-IA* is able to adjust the level of difficulty of the content presented to keep students engaged and motivated to effectively enhance their learning process.<br><br>Here is more relevant information about *Tutor-AI*:<br><br>The level of difficulty of the content presented to the student by *Tutor-AI* [**Agency-High**: is not supervised; **Agency-Low**: is supervised] by the teacher.<br><br>*Tutor-IA* [**Privacy-Low**: shares the personalised learning path with the teacher; **Privacy-High**: does not share the personalised learning path with anyone else]<br><br>*Tutor-IA* [**Explainability-Low**: provides no explanation; **Explainability-High**: provides a detailed explanation] on how personalisation for each student is achieved.<br><br>It [**Transparency-Low**: is not communicated; **Transparency-High:** is clearly communicated] to the student that the level of difficulty of the content presented by *Tutor-IA* is customised for him or her. | Imagine a high school providing students with the *Emotion-AI* system, a chatbot (virtual conversational agent based on artificial intelligence) providing emotional support to students. The student interacts with *Emotion-AI* in the form of a realistic text conversation. The chatbot is able to understand the meaning and context of the conversation and is therefore able to provide students with appropriate emotional support. *Emotion-AI* is able to listen exhaustively to the student and direct them to the appropriate resources if necessary.<br><br>Here is more relevant information about *Emotion-AI*:<br><br>The advice provided by *Emotion-AI* is formulated by an artificial intelligence system [**Agency-High**: without being validated; **Agency-Low**: validated supervised] by educational advisors.<br><br>*Emotion-AI* [**Privacy-Low**: shares the content of the conversation with educational advisors; **Privacy-High**: does not share the content of the conversation with anyone else].<br><br>*Emotion-AI* [**Explainability-Low**: does not explain; **Explainability-High**: explains] its reasoning for the advice provided to the student.<br><br>The student [**Transparency-Low**: is not informed; **Transparency-High:** is informed] that he or she is communicating with a chatbot and not a human. |

# Appendix 2 Modified psychometric scales.

| MEASURES (modified TAM and ETHICS items) | | | |
|---|---|---|---|
| **Contexte 1** | **Contexte 2** | **Contexte 3** | **Contexte 4** |
| Dans quelle mesure êtes-vous d'accord avec les affirmations suivantes concernant l'utilisation de l'IA dans le scénario qui vous a été présenté : | Dans quelle mesure êtes-vous d'accord avec les affirmations suivantes concernant l'utilisation de l'IA dans le scénario qui vous a été présenté : | Dans quelle mesure êtes-vous d'accord avec les affirmations suivantes concernant l'utilisation de l'IA dans le scénario qui vous a été présenté : | Dans quelle mesure êtes-vous d'accord avec les affirmations suivantes concernant l'utilisation de l'IA dans le scénario qui vous a été présenté : |

| **Utilitarisme** | **Utilitarisme** | **Utilitarisme** | **Utilitarisme** |
|---|---|---|---|
| [Utilitarianism1] L'utilisation de cette technologie d'intelligence artificielle de correction des examens mènerait à plus de conséquences positives que de conséquences négatives.<br><br>[Utilitarianism2] L'utilisation de cette technologie d'intelligence artificielle de correction des examens mènerait à une meilleure expérience d'apprentissage pour le plus grand nombre d'élèves.<br><br>[Utilitarianism3] En considérant tous les impacts de cette technologie d'intelligence artificielle de correction des examens, cette technologie serait une bonne chose.<br><br>[Utilitarianism4] En considérant tous les aspects | [Utilitarianism1] L'utilisation de cette technologie d'agent conversationnel basé sur l'intelligence artificielle mènerait à plus de conséquences positives que de conséquences négatives.<br><br>[Utilitarianism2] L'utilisation de cette technologie d'agent conversationnel basé sur l'intelligence artificielle mènerait à une meilleure expérience d'apprentissage pour le plus grand nombre d'élèves.<br><br>[Utilitarianism3] En considérant tous les impacts de cette technologie d'agent conversationnel basé sur l'intelligence artificielle, cette technologie serait une bonne chose.<br><br>[Utilitarianism4] En considérant tous les | [Utilitarianism1] L'utilisation de cette technologie de plateforme personnalisée d'apprentissage basée sur l'intelligence artificielle mènerait à plus de conséquences positives que de conséquences négatives.<br><br>[Utilitarianism2] L'utilisation de cette technologie de plateforme personnalisée d'apprentissage basée sur l'intelligence artificielle mènerait à une meilleure expérience d'apprentissage pour le plus grand nombre d'élèves.<br><br>[Utilitarianism3] En considérant tous les impacts de technologie de plateforme | [Utilitarianism1] L'utilisation de cette technologie d'agent virtuel de soutien émotionnel mènerait à plus de conséquences positives que de conséquences négatives.<br><br>[Utilitarianism2] L'utilisation de cette technologie d'agent virtuel de soutien émotionnel mènerait à une meilleure expérience d'apprentissage pour le plus grand nombre d'élèves.<br><br>[Utilitarianism3] En considérant tous les impacts de cette technologie d'agent virtuel de soutien émotionnel, cette technologie serait une bonne chose.<br><br>[Utilitarianism4] En considérant tous les aspects de cette technologie d'agent virtuel de soutien émotionnel, cette technologie serait efficace. |
| de cette technologie d'intelligence artificielle de correction des examens, cette technologie serait efficace. | aspects de cette technologie d'agent conversationnel basé sur l'intelligence artificielle, cette technologie serait efficace. | personnalisée d'apprentissage basée sur l'intelligence artificielle, cette technologie serait une bonne chose.<br><br>[Utilitarianism4] En considérant tous les aspects de technologie de plateforme personnalisée d'apprentissage basée sur l'intelligence artificielle, cette technologie serait efficace. | |

| Justice | Justice | Justice | Justice |
|---|---|---|---|
| [Justice1] L'utilisation de cette technologie d'intelligence artificielle de correction des examens serait plus juste pour les élèves.<br><br>[Justice2] L'utilisation de cette technologie d'intelligence artificielle de correction des examens serait plus impartiale et équitable pour les élèves.<br><br>[Justice3] L'utilisation de cette technologie d'intelligence artificielle de correction des examens mènerait à une évaluation plus fiable des examens, ce qui serait plus juste pour tous les élèves. | [Justice1] L'utilisation de cette technologie d'agent conversationnel basé sur l'intelligence artificielle serait plus juste pour les élèves.<br><br>[Justice2] L'utilisation de cette technologie d'agent conversationnel basé sur l'intelligence artificielle serait plus impartiale et équitable pour les élèves.<br><br>[Justice3] L'utilisation de cette technologie d'agent conversationnel basé sur l'intelligence artificielle rendrait le soutien à l'apprentissage plus accessible, ce qui serait plus juste pour tous les élèves. | [Justice1] L'utilisation de cette technologie de plateforme personnalisée d'apprentissage basée sur l'intelligence artificielle serait plus juste pour les élèves.<br><br>[Justice2] L'utilisation de cette technologie de plateforme personnalisée d'apprentissage basée sur l'intelligence artificielle serait plus impartiale et équitable pour les élèves.<br><br>[Justice3] L'utilisation de cette technologie de plateforme personnalisée | [Justice1] L'utilisation de cette technologie d'agent virtuel de soutien émotionnel serait plus juste pour les élèves.<br><br>[Justice2] L'utilisation de cette technologie d'agent virtuel de soutien émotionnel serait plus équitable et sans jugement pour les élèves.<br><br>[Justice3] L'utilisation de cette technologie d'agent virtuel rendrait le soutien émotionnel plus accessible, ce qui serait plus juste pour tous les élèves. |

|  |  | d'apprentissage basée sur l'intelligence artificielle rendrait le processus d'apprentissage plus personnalisé, ce qui serait plus équitable pour tous les élèves. |  |
|---|---|---|---|

| Confiance / Trust | Confiance / Trust | Confiance / Trust | Confiance / Trust |
|---|---|---|---|
| [Trust1] J'ai confiance que l'utilisation de cette technologie d'intelligence artificielle de correction des examens fournirait une évaluation fiable de mes apprentissages.

[Trust2] Je suis convaincu(e) que l'utilisation de cette technologie d'intelligence artificielle de correction des examens serait plus précise et fiable que celle d'un enseignant.

[Trust3] Je ferais confiance aux enseignants et à l'école pour assurer la sécurité et la confidentialité de mes données personnelles dans l'utilisation de cette technologie d'intelligence artificielle de correction des examens. | [Trust1] J'ai confiance que l'utilisation de cette technologie d'agent conversationnel basé sur l'intelligence artificielle fournirait des réponses fiables à mes questions.

[Trust2] Je suis convaincu(e) que l'utilisation de cette technologie d'agent conversationnel basé sur l'intelligence artificielle offrirait des réponses plus précises et fiables que celles offertes par un enseignant.

[Trust3] Je ferais confiance aux enseignants et à l'école pour assurer la sécurité et la confidentialité de mes données personnelles dans l'utilisation de cette technologie d'agent conversationnel basé sur l'intelligence artificielle. | [Trust1] J'ai confiance que l'utilisation de cette technologie de plateforme personnalisée d'apprentissage basée sur l'intelligence artificielle fournirait un processus d'apprentissage fiable.

[Trust2] Je suis convaincu(e) que l'utilisation de cette technologie de plateforme personnalisée d'apprentissage basée sur l'intelligence artificielle offrirait un processus d'apprentissage plus précis et fiable que celui offert par un enseignant.

[Trust3] Je ferais confiance aux enseignants et à l'école pour assurer la sécurité et la | [Trust1] J'ai confiance que l'utilisation de cette technologie d'agent virtuel offrirait un soutien émotionnel fiable aux élèves.

[Trust2] Je suis convaincu(e) que l'utilisation de cette technologie d'agent virtuel offrirait un soutien émotionnel plus précis et fiable que celui offert par les enseignants et le personnel scolaire.

[Trust3] Je ferais confiance aux enseignants, conseillers scolaires et à l'école pour assurer la sécurité et la confidentialité de mes données personnelles dans l'utilisation de technologie d'agent virtuel de soutien émotionnel. |

| | | | confidentialité de mes données personnelles dans l'utilisation de cette technologie de plateforme personnalisée d'apprentissage basée sur l'intelligence artificielle. | |

| **Risque / Risk** | **Risque / Risk** | **Risque / Risk** | **Risque / Risk** |
|---|---|---|---|
| [Risk1] Il existe une possibilité de dysfonctionnement et de défaillance des systèmes, de sorte que la technologie d'intelligence artificielle de correction automatique pourrait ne pas être en mesure de fournir une évaluation fiable de mes examens.<br><br>[Risk2] Dans le cadre de l'utilisation de cette technologie de correction automatique des examens basé sur l'intelligence artificielle, je crains que mes données personnelles et d'apprentissage ne soient pas sécurisées et puissent être consultées par des personnes non autorisées, ce qui pourrait entraîner une utilisation abusive de mes données à mon insu et de la discrimination à mon égard.<br><br>[Risk3] Compte tenu de la complexité des systèmes d'intelligence artificielle, je pense qu'il existe un risque que les résultats de mes évaluations soient incorrects. | [Risk1] Il existe une possibilité de dysfonctionnement et de défaillance des systèmes, de sorte que la technologie d'agent conversationnel basé sur l'intelligence artificielle pourrait ne pas être en mesure de fournir des réponses fiables à mes questions.<br><br>[Risk2] Dans le cadre de l'utilisation de cette technologie d'agent conversationnel basé sur l'intelligence artificielle, je crains que mes données personnelles et d'apprentissage ne soient pas sécurisées et puissent être consultées par des personnes non autorisées, ce qui pourrait entraîner une utilisation abusive de mes données à mon insu et de la discrimination à mon égard.<br><br>[Risk3] Compte tenu de la | [Risk1] Il existe une possibilité de dysfonctionnement et de défaillance des systèmes, de sorte que la technologie de plateforme personnalisée d'apprentissage basée sur l'intelligence artificielle pourrait ne pas être en mesure de fournir un processus d'apprentissage personnalisé approprié à ma situation.<br><br>[Risk2] Dans le cadre de l'utilisation de cette technologie de plateforme personnalisée d'apprentissage basée sur l'intelligence artificielle, je crains que mes données personnelles et d'apprentissage ne soient pas sécurisées et puissent être consultées par des parties non autorisées, ce qui | [Risk1] Il existe une possibilité de dysfonctionnement et de défaillance des systèmes, de sorte que la technologie d'agent virtuel de soutien émotionnel pourrait ne pas être en mesure de fournir un soutien émotionnel approprié à ma situation.<br><br>[Risk2] Dans le cadre de l'utilisation de cette technologie d'agent virtuel de soutien émotionnel, je crains que mes données personnelles et émotionnelles ne soient pas sécurisées et puissent être consultées par des personnes non autorisées, ce qui pourrait entraîner une utilisation abusive de mes données à mon insu et de la discrimination à mon égard.<br><br>[Risk3] Compte tenu de la complexité des systèmes d'intelligence artificielle, je pense qu'il existe un risque que le support émotionnel qui me soit offert soit |
| [Risk4] Parce que je pourrais avoir de la difficulté à comprendre correctement le rapport de correction de mes travaux fait par l'IA, cela pourrait augmenter mon anxiété concernant mes études. | complexité des systèmes d'intelligence artificielle, je pense qu'il existe un risque que les réponses à mes questions soient incorrectes.<br><br>[Risk4] Parce que je pourrais avoir de la difficulté à comprendre correctement les réponses données par la technologie d'agent conversationnel basé sur l'intelligence artificielle, cela pourrait augmenter mon anxiété concernant mes études. | pourrait entraîner une utilisation abusive de mes données à mon insu et de la discrimination à mon égard.<br><br>[Risk3] Compte tenu de la complexité des systèmes d'intelligence artificielle, je pense qu'il existe un risque que le processus d'apprentissage personnalisé qui me soit proposé soit incorrect.<br><br>[Risk4] Parce que je pourrais avoir de la difficulté à comprendre correctement le processus d'apprentissage personnalisé proposé par l'IA, cela pourrait augmenter mon anxiété concernant mes études. | incorrect.<br><br>[Risk4] Parce que je pourrais avoir de la difficulté à comprendre correctement le support émotionnel offert par l'IA, cela pourrait augmenter mon anxiété concernant ma situation. |

**Example Stakeholder adjustment Parents**

Utilitarisme Dans quelle mesure êtes-vous d'accord avec les affirmations suivantes concernant l'utilisation de Correct-IA dans le scénario qui vous a été présenté ?

| Utilitarianisme 1 | | | |
|---|---|---|---|
| L'utilisation de Correct-IA mènerait à plus de conséquences positives que de conséquences négatives | L'utilisation de Correct-IA mènerait à une meilleure expérience d'apprentissage pour plus grand nombre d'élèves. | L'utilisation de Correct-IA mènerait à une meilleure experience d'apprentissage seulement pour les élèves les plus performants à l'école. | Pour notre vérification de votre niveau d'attention, veuillez cocher la case " tout à fait en désaccord". |
| **Utilitarianisme 2** | | | |
| L'utilisation de Correct-IA mènerait à une meilleure expérience d'apprentissage seulement pour les élèves avec des difficultés scolaires. | En considérant tous les impacts de Correct-IA, cette technologie serait une bonne chose. | En considérant tous les aspects de Correct-IA, cette technologie serait efficace. | |

**Example Stakeholder adjustment Teachers**

Utilité perçue Dans quelle mesure êtes-vous d'accord avec les affirmations suivantes concernant l'utilisation de Réponse-IA dans le scénario qui vous a été présenté ?

| Utilité 1 | | | |
|---|---|---|---|
| L'utilisation de Réponse-IA aiderait mes élèves avec leurs apprentissages. | L'utilisation de Réponse-IA donnerait à mes élèves des informations détaillées sur leurs apprentissages, ce qui serait très utile. | L'utilisation de Réponse-IA aiderait les enseignants et l'école à mieux identifier les élèves avec des difficultés scolaires. | L'utilisation de Réponse-IA aiderait l'école à réaliser des économies. |
| **Utilité 2** | | | |
| L'utilisation de Réponse-IA serait un bon complément à l'approche actuelle. | L'utilisation de Réponse-IA offrirait de meilleures réponses aux questions de mes élèves que celles offertes seulement par un enseignant. | L'utilisation de Réponse-IA serait plus accessible pour mes élèves que les réponses aux questions offertes seulement par un enseignant. | |